\documentclass[12pt, a4paper]{article}
\usepackage[font=footnotesize]{caption}
\usepackage[utf8]{inputenc}
\usepackage{graphicx}
\usepackage{authblk}
\usepackage{booktabs}
\usepackage{verbatim}
\usepackage{url}
\usepackage{caption}
\usepackage{subcaption}
\usepackage{indentfirst}
\usepackage{url}
\usepackage{amssymb}
\usepackage{amsmath}
\title{Simple Bundles of Complex Networks}
\author{Alexandre Benatti$^1$ and Luciano da F. Costa$^2$}

\affil{
$^1$Institute of Mathematics and Statistics - DCC \\
University of S\~ao Paulo \\
Rua do Mat\~ao, 1010, S\~ao Paulo, SP 05508-090 Brazil %\\ \vspace{0.5cm}
\\ \vspace{0.5cm}
$^2$S\~ao Carlos Institute of Physics - DFCM \\
University of S\~ao Paulo \\
Av. Trabalhador S\~ao-Carlense, 400, S\~ao Carlos, SP 13566-590 Brazil
}

\date{\today}

\begin{document}

\maketitle

\begin{abstract}
Complex networks can be used to represent and model an ample diversity of abstract and real-world systems and structures. A good deal of the research on these structures has focused on specific topological properties, including node degree, shortest paths, and modularity. In the present work, we develop an approach aimed at identifying and characterizing simple bundles of interconnections between pairs of nodes (source and destination) in complex networks. More specifically, simple bundles can be understood as corresponding to the bundle of paths obtained while traveling through successive neighborhoods after departing from a given source node. Because no node appears more than once along a given bundle, these structures have been said to be simple, in analogy to the concept of a simple path. In addition to describing simple bundles and providing a possible methodology for their identification, we also consider how their respective effective width can be estimated in terms of diffusion flow and exponential entropy of transition probabilities. The potential of the concepts and methods described in this work is then illustrated respectively to the characterization and analysis of model-theoretic networks, with several interesting results.
\end{abstract}

\section{Introduction}

We live in an age characterized by increasing interconnectivity at almost every level and aspect of human experience and activity, from the Internet to cultural aspects. Interestingly, the interconnectivity of real-world systems not only tends to increase with time, but this often takes place in terms of distinct patterns of connections which can have a critical impact on several related properties as well as dynamics taking place in those systems.

As a consequence of their intrinsic ability to represent virtually any discrete interconnected system, complex networks have achieved particular theoretical and applied importance, constituting the main subject studied in the new area of \emph{network science} (e.g.~\cite{barabasi2013network}).

Once an interconnected structure or system has been represented as a complex network, it becomes possible to characterize several of its topological properties in terms of respective measurements (e.g.~\cite{costa2007characterization}), including but not limited to node degree, clustering coefficient, shortest path lengths, etc.  These measurements can extend along successive topological scales, being respective to each individual node (e.g.~node degree, clustering coefficient), pairs of nodes (e.g.~assortativity and shortest distance), subgraphs (e.g.~size and statistics of respective topological features), and entire networks (e.g.~size and statistics of respective topological features).

Given a network and two respective distinct nodes, one particularly interesting property of the interconnections between this pair of nodes consists of the distribution of \emph{simple paths} between them. Basically, a simple path is a path along which no node is repeated.  This type of subgraph is of particular importance because it implements a more straightforward connection between the two nodes, without the redundant loops that would be otherwise implied by repeated nodes.

\begin{figure}
  \centering
     \includegraphics[width=0.55\textwidth]{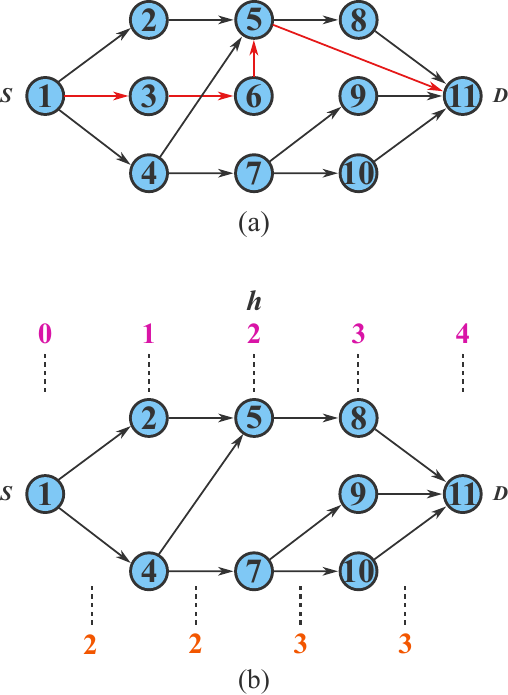}
   \caption{(a): Illustration of the subgraph of a given network (not shown) corresponding to the set of simple paths with length $L=4$ comprised between two nodes (1 and 11). The source and destination nodes have been abbreviated as $S$ and $D$. Observe that one of the simple paths, shown in red, has more than one node at neighborhood hierarchy $h=2$. This type of path precludes the determination of the number of links at each hierarchy. (b): The simple bundle obtained respectively to the same network as considered in (a) corresponds to one of its possible subgraphs. Given that all successive nodes now are in the same hierarchies, it is now possible to determine the respective number of links, shown underneath each hierarchical level.  The concept of \emph{effective width} of the simple bundle, addressed in the present work, allows a more objective characterization of the flow efficiency from source to destination, therefore providing an indication about the independence between the respective constituent paths.}\label{fig:simple}
\end{figure}

Because the simple paths between two nodes usually have several lengths, it becomes important to perform respective analyses for each of those lengths $L$, leading to a multi-scale methodology. Several simple paths will be typically observed for a given value of $L$, as illustrated in Figure~\ref{fig:simple}(a).

One of the limitations of subgraphs such as that illustrated in Figure~\ref{fig:simple}(a) is that two successive nodes along a simple path may result in the same hierarchy, therefore undermining the objective definition of a system of hierarchies emanating from the source node, which is necessary in order that the number of links, as well as the respective effective width at each hierarchy, cannot be specified.

In order to avoid the above observed limitation of using simple paths, we resource to the concept of \emph{simple bundles} comprised between a pair of nodes. Figure~\ref{fig:simple}(b) illustrates the simple bundle obtained for the same network from which the simple paths subgraph in (a) has been obtained. Simple bundles have the important property that each successive node along the several paths comprised between the source and destination nodes belongs to the same hierarchical level $h$. This property then allows the effective width of the bundle to be objectively determined along each hierarchical level, shown underneath the simple bundle in Figure~\ref{fig:simple}(b). The identification of the simple bundles between a pair of network nodes is described in Section~\ref{sec:simplebundles} of the present work, which also introduces the concept of \emph{effective bundle width} quantifying how much the involved paths are independent of one another.

It should be observed that, though the above example referred to a single source and a single destination node, simple bundles can also be defined respectively to any number of source and destination nodes. However, for simplicity's sake, the present work focuses on simple bundles defined respectively to single pair of source and destination nodes.

The connectivity and dependence of the simple paths linking network nodes can indicate their effectiveness across several domains, such as (a) resilience against attacks, (b) achieving the maximal flow, and (c) implied costs (number of links and nodes along the paths).

It is important to observe that the analysis of the interconnection between two network nodes in terms of the respectively comprised simple paths or simple bundles involves selecting these specific subgraphs while leaving out the remainder nodes and links in the analyzed network. This type of approach has been adopted in various other topological analyses of networks. For example, the concept of node degree only takes into account the links that come from that particular node, regardless of the rest of the network. Similarly, the shortest distance between a pair of nodes considers only the respective shortest path(s), leaving out all other network nodes and links. However, each of these methods needs to be motivated and justified.

In the specific case of the simple bundles considered in the present work, there are several reasons for considering this type of subgraph. First, as already discussed, it leads to a well-defined system of hierarchies starting at the source node, which therefore allows the width of the bundle (and respective efficiency) to be objectively determined for each of these hierarchies, which can then be summarized in terms of the respective averages and standard deviations. In addition, the simple bundle between a pair of network nodes will typically have fewer interconnections than the set of respective simple paths, therefore keeping the number of possible combinations at a more manageable level.

More practical justifications for using simple bundles relate to several types of problems which are intrinsically related to this type of structure. For instance, we have dynamics such as a progressive dilation taking place from the source node, in which case the neighboring nodes are reached (or covered) successively and without repetition, in the same manner as the respective simple bundle originating at the same source. This type of dynamics is therefore characteristically found in coverage problems.

In situations as that illustrated in Figure~\ref{fig:simple}, where the original network involves additional paths between a given pair of source-destination nodes, the study of the respective simple bundles remains interesting because the latter type of structure is necessarily contained in the original network.  Therefore, only the links corresponding to a simple bundle can be taken into account while implementing specific dynamics between the respective source-destination nodes.     

The present work aims at characterizing a given network, or a set of its subgraphs, in terms of the concept of effective width~\cite{da2023quantifying}, while considering some specific or all the possible pairs of distinct nodes as source and destination of respective simple bundles.  

The main motivation for this study consists in the fact that these measurements can provide a valuable indication of how well the pairs of nodes in a given network are interconnected from the perspective of simple bundles having lengths of specific interest. For instance, it may be found that a given network has well-interconnected pairs for a given length, decreasing for other lengths. 

Well-interconnected networks or subgraphs are of particular interest due to their ability to facilitate efficient flow between the source and destination nodes. This indicates that the paths in simple bundles operate independently, enhancing the network's resilience to node attacks. Simple bundles that rely on a convergence of individual paths can create bottlenecks and are less robust than those with good flow connections.

Therefore, the characterization of the simple bundles of a network has great potential for applications in several areas and types of networks, including transportation networks (airports and highways), urban networks, word and concept associations, as well as economic interactions.

This work starts by presenting the adopted basic concepts, which include the definition of simple paths, simple bundles, entropy, and exponential entropy. Next, the method employed for finding the equilibrium flow along simple bundles is described and illustrated. The potential of the described concepts and methods is then illustrated respectively to perfect as well as geometrically and topologically modified geographical networks (Delaunay triangulations).

\section{Related Works}\label{sec:relatedworks}

The present section provides a brief review of some of the works related to the developments reported in the current work, including types of topological measurements, hierarchical neighborhoods, accessibility, as well as measurements related to pairs of nodes.

While the degree of the nodes in complex networks provides a particularly important characterization of the respective local interconnectivity, it is not enough for fully describing the respective topology (e.g.~\cite{da2018complex}). Additional measurements (e.g.~\cite{costa2007characterization,trusina2004hierarchy,newman2018networks}) are thus required, which may include local properties such as the clustering coefficient (e.g.~\cite{watts1998collective}), interconnectivity features respective to pairs of nodes such as the shortest distance (e.g.~\cite{dijkstra1959note,costa2007characterization,newman2018networks}), as well as more global measurements including number and size of modules or communities (e.g.~\cite{newman2006modularity, newman2012communities}).

One particularly interesting type of topological measure that can be used to characterize complex networks involves \emph{mesoscopic} approaches defined respectively to a reference node, characterized by occupying an intermediate position between local and global measurements. In particular, given a specific node of a network, it is possible to identify the nodes at successively distant neighborhoods, therefore defining a respective system of hierarchical neighborhoods (e.g.~\cite{trusina2004hierarchy,da2006hierarchical, newman2003ego}). In this manner, \emph{signatures} of specific measurements can be obtained respectively to each successive neighborhood. For instance, given a reference node, a signature can be obtained containing the average degrees of the nodes at each of the respective defined neighborhoods. Related approaches have been developed in several works, including but not limited to~\cite{da2022autorrelation,tokuda2023cross}.

A hierarchical approach to characterizing the topology of networks that takes into account not only specific types of dynamics has been described~\cite{travenccolo2008accessibility, benatti2021accessibility}, in which the transition probabilities implied by the considered dynamics are used to estimate the effective interconnection between a given reference node and the nodes at successive respective neighborhoods. For instance, in the case of traditional, non-preferential diffusion on a network, the transition probabilities from a given node can be considered for obtaining the exponential of the entropy of those probabilities considering each successive hierarchical neighborhood, which has been understood as a measurement (called \emph{accessibility}) of how much that reference node can interact with the nodes at each considered neighborhood. Thus, maximum interaction is achieved when identical transition probabilities are implied by the considered dynamics between the reference node and the nodes at a given neighborhood. In this case, the accessibility becomes identical to the number of neighbors. The number of paths between two network nodes and the accessibility of a node has been considered in~\cite{rodrigues2009structure} as a means to characterize pairwise node interaction in complex networks.

In another interesting type of approach, topological measurements are obtained respectively to \emph{pairs of nodes} of a given network, which will be henceforth referred to as the \emph{pairwise approach}. Prototypical examples of this type of measurement include the shortest path (e.g.~\cite{dijkstra1959note,costa2007characterization,newman2018networks}), as well as the matching index (e.g.~\cite{kaiser2004edge}), between a pair of network nodes. Shortest paths have received deserved attention because of their relevance in characterizing and influencing several aspects not only of the network structure (e.g.~\cite{gloaguen2010analysis,domingues2020shortest,de2016minimal}) but also of dynamics taking place in those networks.

The present work combines the above revised hierarchical and pairwise approaches to topological measurements characterizing individual nodes in a complex network.

\section{Basic Concepts}\label{sec:concepts}

Given an undirected graph, or complex network, and two respective distinct nodes $a$ and $b$, a \emph{simple path} of length $L$ extending between these two nodes will consist of a sequence of adjacent links starting passing through $L-1$ distinct nodes (not considering $a$ and $b$). In the present work, we shall consider paths extending from a source to a destination node. The nodes of the paths between the source and destination are henceforth said to be \emph{intermediate}.

As discussed in the Introduction, the subgraph formed by the simple paths between two network nodes may lack a clear neighborhood hierarchy starting from the source node. That happens when one or more nodes appear at more than one distance from the source.

In order to ensure a well-defined hierarchy of neighborhoods, we shall focus on \emph{simple bundles} comprised between a source and a destination node. These bundles correspond to the set of paths obtained while considering paths proceeding from each successive neighborhood around the source node.  This is illustrated in Figure~\ref{fig:method} respectively to a small network.

Given a set of discrete probabilities $p_i$, $i = 1, 2, \ldots, N$, normalized so that:
\begin{align}
  \sum_{i=1}^N p_i = 1
\end{align}

the respective \emph{entropy} (e.g.~\cite{cover2006elements}) can be defined as:
\begin{align}
  \varepsilon = \sum_{i=1}^N p_i \, \log(p_i)
\end{align}

The respective \emph{exponential entropy} (e.g.~\cite{campbell1966exponential,pielou1966shannon,travenccolo2008accessibility,benatti2022complex,viana2012effective}) can thus be defined as corresponding to:
\begin{align}
  \eta = e^{\varepsilon} 
\end{align}

It can be shown that this quantity is lower bounded by 0 and upper bounded by $N$. The maximum bound for a given $N$ is achieved whenever $p_i = 1/N$.

The exponential entropy can be argued to provide an estimation of the \emph{effective number of choices} implemented by the set of probabilities $p_i$. For instance, no choices ($\eta=0$) happen whenever $N=1$ and $p_1=1$. The maximum number of choices $\eta=N$ is observed for $p_i=1/N$. Intermediate situations are defined for $0 < \eta < 1$.

\section{Methodology}\label{sec:methodology}

This section describes the methodologies adopted for identifying the simple bundles between a given pair of source-destination nodes, as well as the estimation of the effective width of simple bundles by considering the entropy of the respective flow at equilibrium.

\subsection{Identifying Simple Bundles}\label{sec:simplebundles}

Before a given complex network can be characterized in terms of the properties of its respective simple bundles, it is first necessary to identify these structures. A respective methodology is described in the following. All networks are henceforth assumed to be originally undirected.

\begin{figure}
  \centering
     \includegraphics[width=0.9\textwidth]{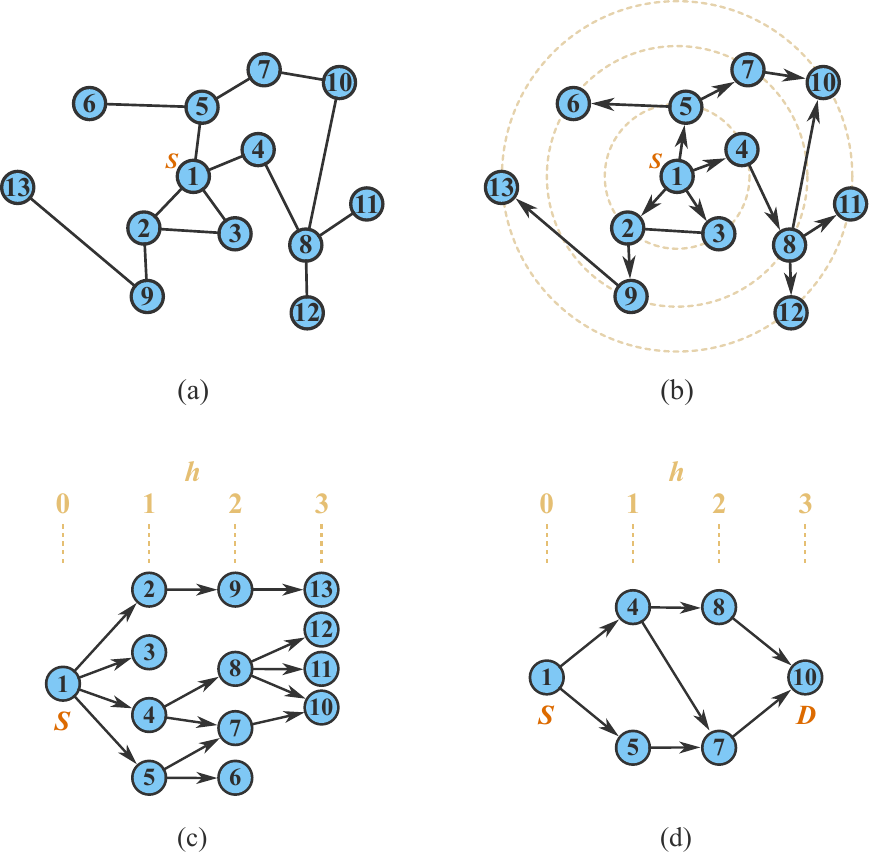}
   \caption{(a): A small network and the three neighborhoods respectively defined around the source node 1. (b): The directed links established between each successive neighborhood. (c): The same structure as in (b), but with the hierarchical neighborhoods organized as respective columns and removal of the links between nodes in the same hierarchy. The simple paths extending from the source node up to each of the identified destination nodes (10, 11, 12, and 13) can be identified by following all possible output links from each node, yielding 8 simple paths. (d): The simple bundle identified between nodes 1 and 10 comprises 3 simple paths: $1 \rightarrow 4 \rightarrow 8 \rightarrow 10$, $1 \rightarrow 4 \rightarrow 7 \rightarrow 10$ and
   $1 \rightarrow 5 \rightarrow 7 \rightarrow 10$.}\label{fig:method}
\end{figure}

Given a network and one of its nodes selected as \emph{source}, a respective system of hierarchical neighborhoods can be established by taking into account the network nodes that are at topological distances $h = 1, 2, \ldots$ from the source node. Figure~\ref{fig:method}a illustrates a small network, with node 1 chosen as the source, as well as its three successive hierarchical neighborhoods, identified by respective dashed circles. The source node 1 can be understood as belonging to the 0-th neighborhood. The nodes at the first neighborhood ($h=1$) are 2, 3, 4, and 5. Those in the second neighborhood ($h=2$) include nodes 6, 7, 8, and 9. The last neighborhood ($h=3$) contains nodes 10, 11, 12, and 13.

Figure~\ref{fig:method}b illustrates the same network as in (a), but with directed links between successive hierarchical neighborhoods. This same structure, but with the hierarchical neighborhoods represented as columns, and with the links between nodes belonging to the same hierarchical neighborhood having been removed, is shown in (c).

Starting at the source node, simple paths can be identified by following all possible outgoing links, one at a time, until all links have been covered. This can be implemented recursively, with the help of a stack data structure. 

While at a node with two or more outgoing links, one of these links is taken as a continuation of the current path, while the other links are pushed into the stack. Once a node with no outgoing link is reached, it is taken as a possible destination of the given source node, and the next path is continued from the link popped out from the stack. This procedure continues until all links have been covered.

The simple bundle defined between the source and each of the possible destination nodes can then be obtained by joining the preliminary identified simple paths extending between those two nodes. For instance, Figure~\ref{fig:method}d depicts the simple bundle identified between nodes 1 and 10, which comprises the paths $1 \rightarrow 4 \rightarrow 8 \rightarrow 10$, $1 \rightarrow 4 \rightarrow 7 \rightarrow 10$ and $1 \rightarrow 5 \rightarrow 7 \rightarrow 10$.

\begin{figure}
  \centering
     \includegraphics[width=0.7\textwidth]{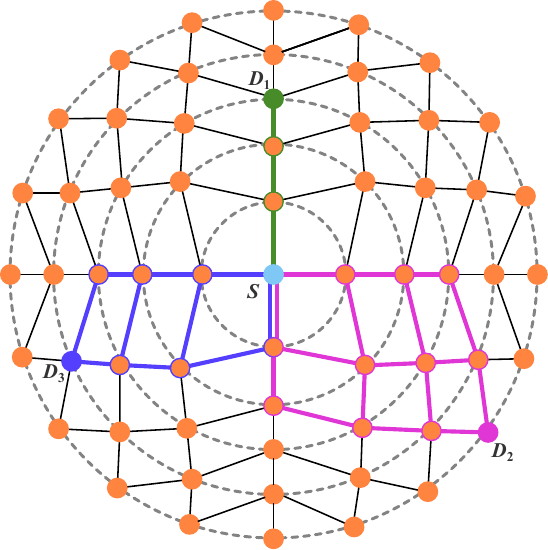}
   \caption{The hierarchical neighborhoods defined onto an orthogonal lattice respectively to taking its most central node as a source. The simple bundles defined by the destination nodes $D_1$, $D_2$, and $D_3$ for lengths $L=3, 5, 4$ are shown in green, magenta, and blue. Each of these bundles contains 1, 10, and 4 constituent paths, respectively.}\label{fig:orthog}
\end{figure}

Figure~\ref{fig:orthog} illustrates three simple bundles defined by respective destination nodes $D_1$, $D_2$, and $D_3$, with the source node placed at the center of an orthogonal lattice dimension $5 \times 5$.

\subsection{Estimating the Effective Width}\label{sec:effectivewidth}

In this section, we will define the concept of effective width of a simple bundle, and show how to calculate this measurement.

\begin{figure}
  \centering
     \includegraphics[width=0.55\textwidth]{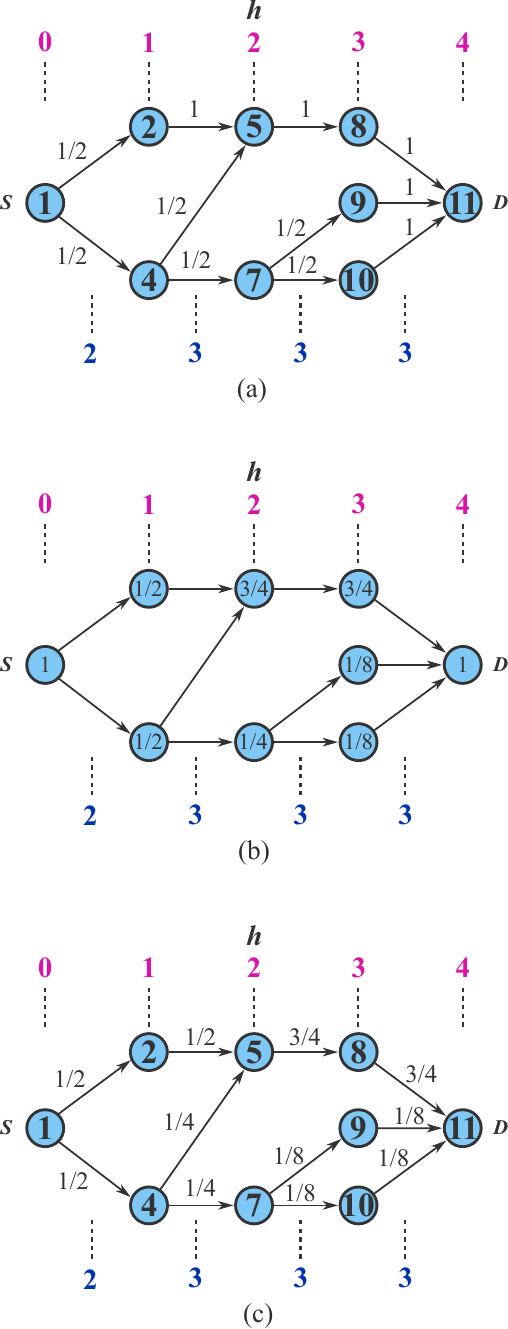}
   \caption{Example of calculation of the equilibrium flow respectively to the simple bundle in Fig.~\ref{fig:method}. (a): The transition probability at each link. (b): The nodes flow at the equilibrium state. (c): The equilibrium flow at each link. The numbers of links per
   hierarchical level are shown (in dark blue) underneath each simple bundle.}\label{fig:flow}
\end{figure}

An illustrative example of how to determine the equilibrium flow rate at each link is shown in Figure~\ref{fig:flow}. The first step involves determining the transition probability for each link of the considered simple bundle.

This can be done by considering each node, assigning to each of the $k$ outgoing links $1/k$ as the respective transition probability, yielding the results shown in Figure~\ref{fig:flow}(a). These transition probabilities can be organized as a transition matrix $T$, in which $T_{ij}$ represents the transition probability from node $i$ to $j$.

The flow of nodes at the equilibrium state can be obtained from the transition probabilities, as illustrated in Figure~\ref{fig:flow}(b). To achieve this, the sum of the incoming transitions to each node must be calculated. The vector with the equilibrium state flow of each node can be determined by the following equations:

\begin{equation}\label{eq:flowh}
    \vec{\phi}_h = T \, \vec{\phi}_{h-1},
\end{equation}

\begin{equation}\label{eq:flow}
    \vec{\phi} = \sum_{h=0}^L \vec{\phi}_{h},
\end{equation}
where $L$ is the bundle lengths and $\vec{\phi_0}$ represents a vector in which all entries are zero, except for the source node, whose value is one.

The node flow $\vec{\phi}$ and transition matrix $T$ can be used to calculate the equilibrium flow at each link, see Figure~\ref{fig:flow}(c). The equilibrium flow of the link between nodes $i$ and $j$, $w_{i,j}$, is denoted by:

\begin{equation}\label{eq:equilibriumlink}
    w_{i,j} = \phi_jT_{ij},
\end{equation}

%-caminhos efetimos entre uma camada e outra
The effective width between the $h-1$ and $h$ neighborhoods of a given simple bundle can be calculated using the equilibrium flow as follows:

\begin{equation}\label{eq:EffectiveWidth}
    E_h = \exp\left(- \sum_{k \in H} \omega_k \log(\omega_k)\right),
\end{equation}
where $H$ is the set of edges between $h-1$ and $h$.

Though the set of values $E_h$ provides a more comprehensive characterization of the effective width of a given simple bundle, therefore also quantifying the independence (and, thus, entanglements) of the constituent paths, it is often interesting to resource to a summarization of these values. Here we use the average of $E_h$ as an overall indication of the width of specific simple bundles, which can be expressed as:
\begin{equation}\label{eq:averageEffectiveWidth}
    \mu_E = \frac{1}{L} \sum_{h=1}^L E_h,
\end{equation}

\begin{figure}
  \centering
     \includegraphics[width=0.55\textwidth]{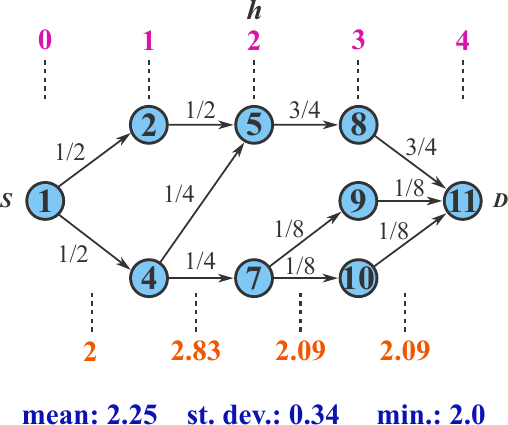}
   \caption{The effective widths obtained for each of the successive hierarchical levels respectively to the simple bundle shown in Fig.~\ref{fig:flow} are shown in orange underneath each hierarchical level. The summarization of the effective widths properties of this simple bundle can be obtained, among other possibilities, in terms of the respective mean, standard deviation, and minimum values.}\label{fig:effWid}
\end{figure}

Figure~\ref{fig:effWid} presents the effective widths obtained for the same simple bundle as in Figure~\ref{fig:flow}, also including the respective summarization in terms of mean, standard deviation, and minimum values.  Observe that the effective widths are always smaller or equal to the respective number of links for the same hierarchical level.

It is interesting to observe that asymmetric links, such as that going from node 4 to node 5, tend to unbalance the subsequent probability flow. In this particular case, became concentrated along the path $5 \rightarrow 8 \rightarrow 11$. An unbalanced flow then implies the effective width to become smaller than the respective number of links at each respective hierarchical level.

\section{Characterizing Model-Theoretic Networks}\label{sec:characterizing}

In this section, we apply the described simple bundle approach to characterize three types of model-theoretic complex networks, namely: (a) two-dimensional, non-periodical orthogonal lattice; (b) same as before, but with node positions displaced by random perturbations; and (c) Watts-Strogatz having the same orthogonal lattice as in (a) as its initial configuration.

Other classical model-theoretical networks, including Erd\H{o}s-R\'enyi and Barab\'asi-Albert graphs, have not been included because of their small diameter, which severely constrains the possible values of $L$ to be considered in the respective characterization and analysis.

We start by considering bundle lengths of $L=2, 3, \ldots, 7$ in the orthogonal lattice with $15 \times 15$ points (and nodes). The lattice central point was chosen as the source, which explains the choice of the lattice dimension to avoid border effects.

\begin{figure}
  \centering
     \includegraphics[width=0.8\textwidth]{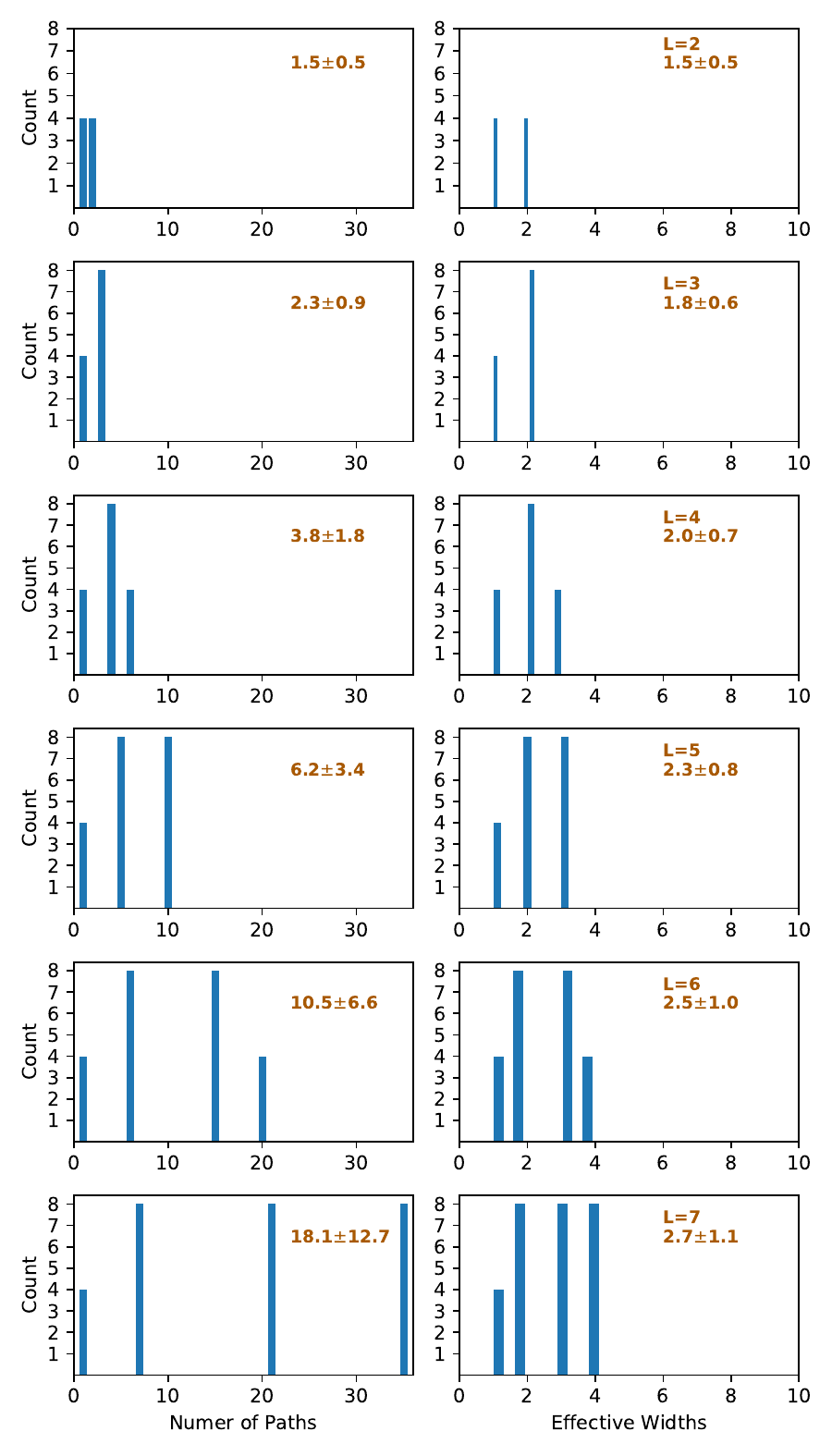}
   \caption{The number of paths (left) and respective effective bundle widths (right) obtained for a $15 \times 15$ lattice, considering its center as the source node, obtained respectively to $L=2, 3, \ldots, 7$. The average $\pm$ standard deviations are respectively indicated. Observe that, in all situations, the distribution of effective widths involves values that are smaller or equal to the respective number of paths. This indicates that the paths composing the simple bundles are highly entangled, therefore reducing the maximum flow and robustness of the simple bundle interconnection between pairs of nodes in the orthogonal lattice.}\label{fig:hist_grid}
\end{figure}

All simple bundles starting at the source node were identified by using the methodology described in Section~\ref{sec:simplebundles}, and each of these bundles was then characterized in terms of the number of paths and respective effective bundle width. The results are shown in the left and right-hand columns in Figure~\ref{fig:hist_grid}.

Several interesting results can be identified from Figure~\ref{fig:hist_grid}. First, observe that the radial symmetry of the lattice implies in respective repetition of the obtained types of simple bundles. In addition, we have that the simple bundles effective widths are invariably smaller than the respective number of paths. This is a direct consequence of the large inter-dependence, or \emph{entanglement} between the respective constituent individual paths in the analyzed bundles, leading to reduced flow and robustness.

Interestingly, the maximum effective width obtained for each considered $L$ increases rather slowly when compared with the respective maximum number of paths, indicating a marked reduction of interconnection efficiency obtained for increasing values of $L$.

All in all, we have that the simple bundles obtained for orthogonal lattices tend to have a maximum number of paths that increases quickly with $L$ while being characterized by intense entanglement between the paths composing the bundles that leads to substantially smaller effective widths, indicating rather limited bundle interconnection efficiency for this type of model-theoretic network.

Now, we proceed to study the effect of adding uniformly distributed spatial perturbations to both the $x$ and $y$ coordinates of the points in the structure considered in the perfect orthogonal lattice considered in the previous example described above.

\begin{figure}
  \centering
     \includegraphics[width=0.8\textwidth]{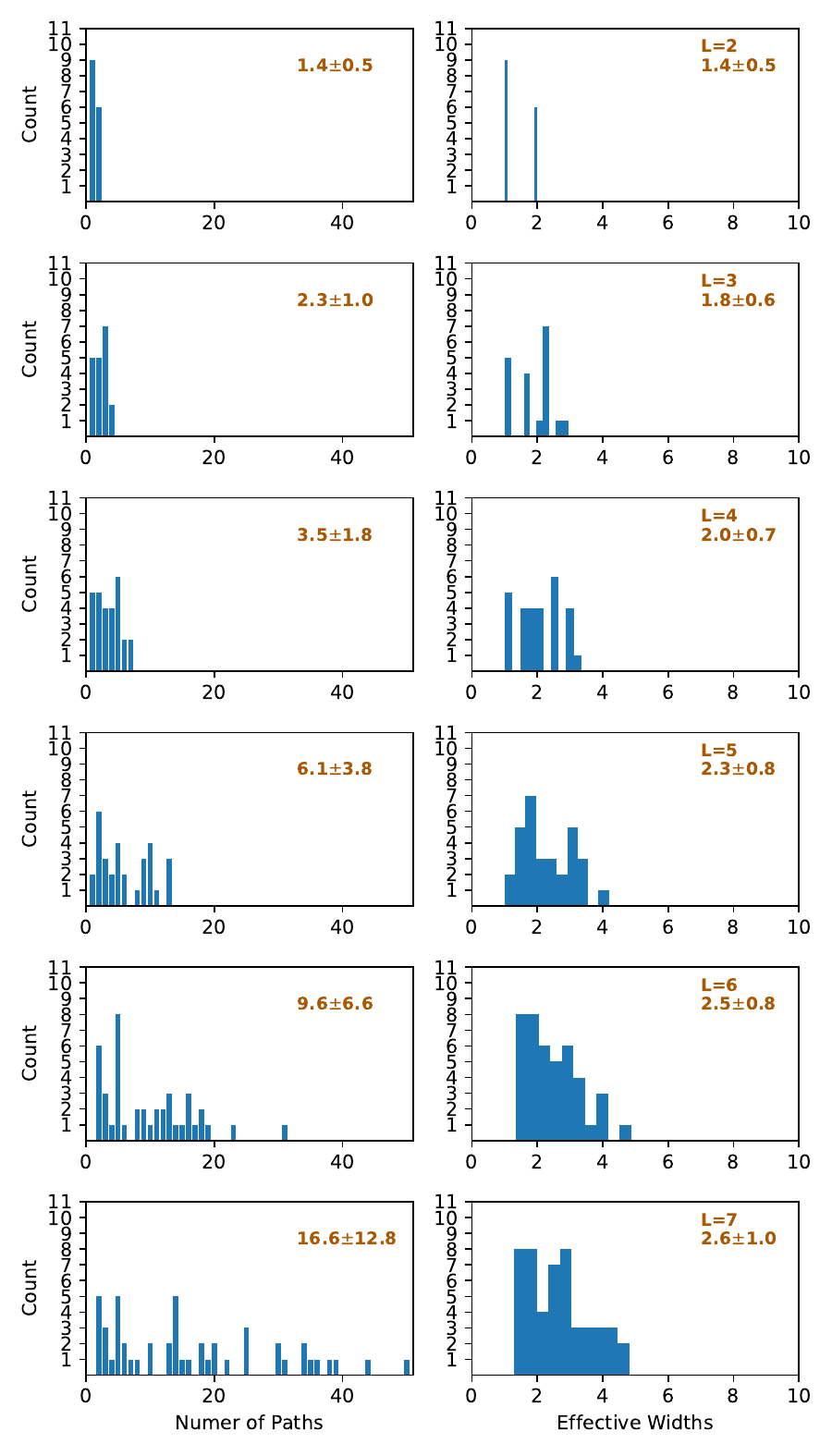}
   \caption{Number of paths and effective widths of the simple bundles identified in the same orthogonal lattice as before, but incorporating uniformly distributed random spatial perturbations $-0.1 \leq \delta \leq 0.1$ imposed onto both the $x$ and $y$ coordinates. Though the number of paths is comparable to those observed for the smaller perturbation case, markedly smaller values of effective widths have been obtained.}\label{fig:hist_voronoi1}
\end{figure}

The number of paths and effective widths obtained for the same orthogonal lattice as before perturbed by $-1e-10 \leq \delta \leq 1e-10$ are presented in the left and right-hand columns of Figure~\ref{fig:hist_voronoi1}.

\begin{figure}
  \centering
     \includegraphics[width=0.8\textwidth]{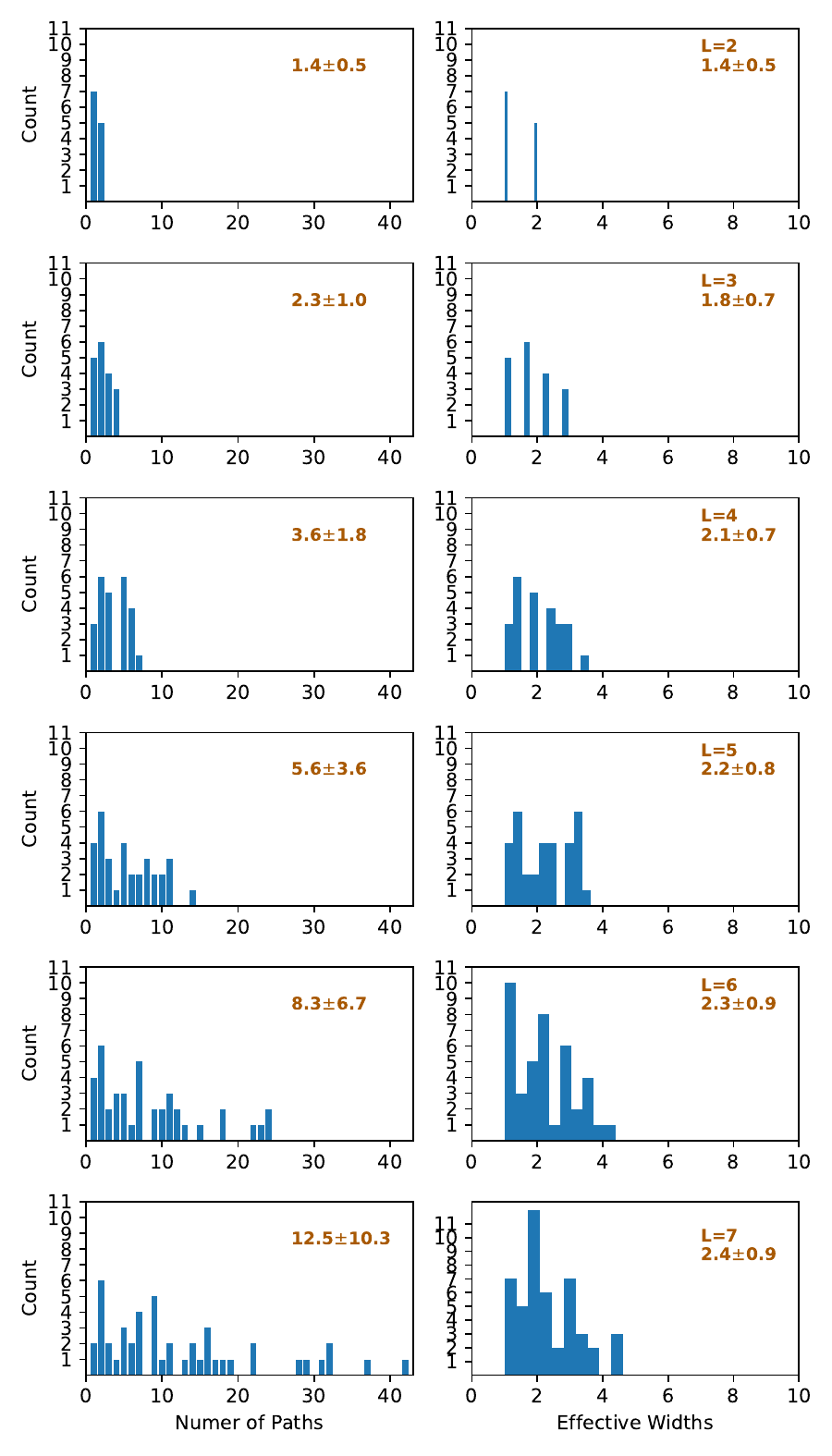}
   \caption{Number of paths and effective widths of the simple bundles identified in the same orthogonal lattice as before, but incorporating uniformly distributed random perturbations $-1e-10 \leq \delta \leq 1e-10$ imposed onto both the $x$ and $y$ coordinates. Remarkably, despite the quite small magnitude of the spatial perturbations $\delta$, the obtained distributions of the number of paths and effective widths results are completely distinct from those obtained previously and shown in Fig.~\ref{fig:hist_grid}. Generally, though the number of paths tended to increase relatively to the previous results, the obtained effective widths are still comparable to those depicted in Fig.~\ref{fig:hist_grid}.}
   \label{fig:hist_voronoi10}
\end{figure}

Surprisingly, markedly different results have been obtained despite the quite small perturbation $\delta$. This is a consequence of the fact that the neighborhoods established in the respective Voronoi tesselation~\cite{lejeune1850reduction,riedinger1988delaunay} (and Delaunay triangulation) can be greatly changed even for small displacements of the respective points.

In addition, though a larger number of paths have been obtained, especially for relatively larger values of $L$, effective widths have been respectively obtained that are mostly comparable to those observed for the perfect orthogonal lattice. 

Figure~\ref{fig:hist_voronoi1} illustrates the number of paths and effective width of the bundles identified for $L=2, 3, \ldots, 7$ respectively to a much larger uniformly random perturbation $-0.1 \leq \delta \leq 0.1$.

Though the resulting number of paths is comparable to those obtained for the smaller perturbation, substantially smaller values of effective widths can be observed. This result indicates that the intensity of the spatial perturbations can strongly influence the interconnection efficiency of the simple bundles.

The results presented and discussed above can be understood as being related to \emph{geometrical} perturbations of the orthogonal lattice. To complement our analysis of model-theoretical networks, we now consider \emph{topological} perturbations to the orthogonal lattice as implemented in the Watts-Strogatz (WS) model starting with that specific configuration.

\begin{figure}
  \centering
     \includegraphics[width=0.8\textwidth]{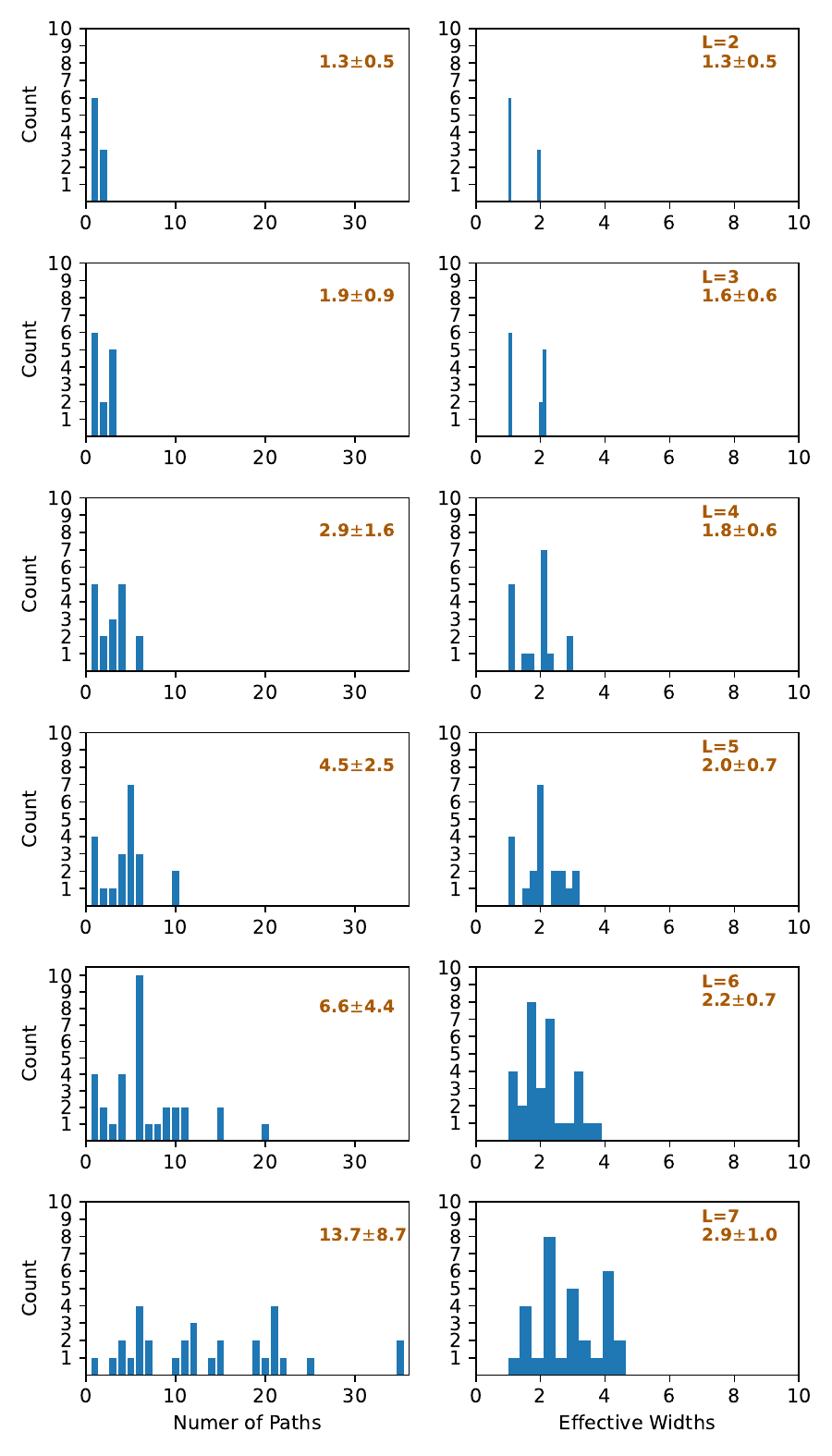}
   \caption{Number of paths and effective widths of the simple bundles identified in the same orthogonal lattice as before, but incorporating rewirings with probability $p=0.02$. Both the number of paths and effective widths are comparable to those obtained for geometrical perturbations using $-0.1 \leq \delta \leq 0.1$ (Fig~\ref{fig:hist_voronoi10}).}\label{fig:hist_WS02}
\end{figure}

Figure~\ref{fig:hist_WS02} shows the number of paths and effective widths obtained for a WS network with the same size as in the previous experiments but subjected to a topological rewiring with probability $p=0.02$.

Interestingly, though the type of perturbation implemented in the latter experiment is completely different from the geometrical perturbations incorporated in the previous cases, the effects on the number of paths and effective width can be observed to be comparable to those cases.

As a general summary of the findings derived from the experiments reported and discussed in the present section, we have that: (a) in all situations the effective width resulted substantially smaller than the respective number of paths; (b) the incorporation of geometrical perturbations tended to relatively increase the number of paths while leading to comparable effective widths; and (c) the incorporation of topological perturbations led to simple bundles with both number of paths and effective width comparable to those obtained for geometrical perturbations with $-0.1 \leq \delta \leq 0.1$. In addition, the effective width values obtained considering all situations addressed in the present section seldom resulted in values larger than $5.0$.

\section{Simple Bundles Networks}  \label{sec:sbn} 

Given a graph/complex network, respective visualizations can be obtained reflecting the effective widths of simple bundles of specific lengths $L$. More specifically, given a network $\Gamma$, a new network $\Lambda$ (henceforth called \emph{simple bundles network}) can be obtained by connecting each of all the possible pairs of distinct nodes in $\Gamma$ with a link having weight corresponding to the average of the average effective widths respective to the two possible simple bundles defined by each pair of nodes.  It is also possible to adopt another measurement --- such as standard deviation,  minimum, or maximum --- instead of the average effective width.

Simple bundles networks therefore provide an effective characterization and visualization of the interactions between pairs of nodes in a given network as expressed by the average effective width of each of the possible simple bundles.  

\begin{figure}
  \centering
     \includegraphics[width=1\textwidth]{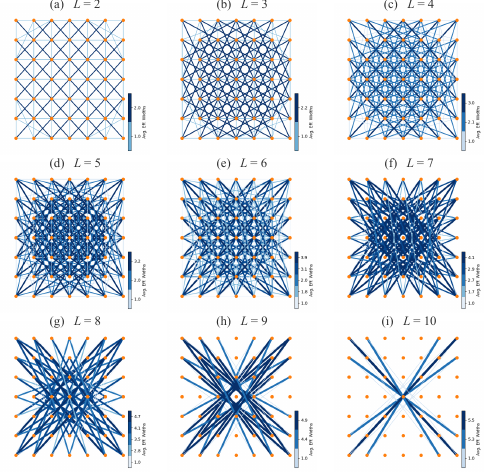}
   \caption{Simple bundles network for $L = 2, 3, \ldots, 10$ of a non-periodical orthogonal lattice with dimension $7 \times 7$. As could be expected, all results present intrinsic symmetry implied by the perfect orthogonal lattice. Observe also that, as $L$ increases, the simple bundles tend to concentrate between nodes at the border of the lattice.}\label{fig:Widths_L}
\end{figure}

Figure~\ref{fig:Widths_L} presents the simple bundles networks for values of $L= 2, 3, \ldots, 10$ of a non-periodical open orthogonal lattice with dimension $7 \times 7$.  As shown in Figure~\ref{fig:Widths_L}(b), the simple bundles for $L=3$ have only two average effective widths: 1.0 and 2.2. The obtained networks present a marked symmetry implied by the symmetry of the original lattice.

\begin{figure}
  \centering
     \includegraphics[width=.49\textwidth]{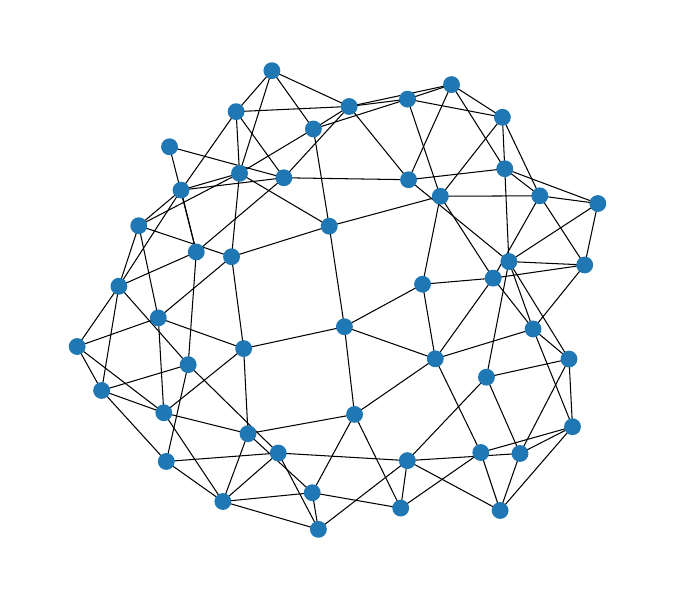}
     \includegraphics[width=.49\textwidth]{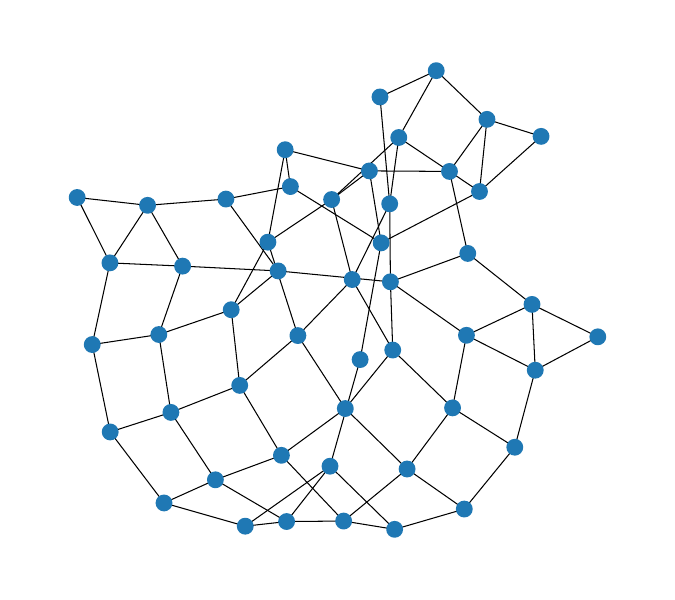}
     \\ \vspace{.1cm} (a) \hspace{.4\textwidth} (b)
   \caption{Visualization, by using the Kamada-Kawai~\cite{kamada1989algorithm} methodology, of the Delaunay network obtained for the non-periodical orthogonal lattice modifications of the node positions (a) and the Watts-Strogatz networks rewired with $p=0.05$.}\label{fig:nets}
\end{figure}

\begin{figure}
  \centering
     \includegraphics[width=0.7\textwidth]{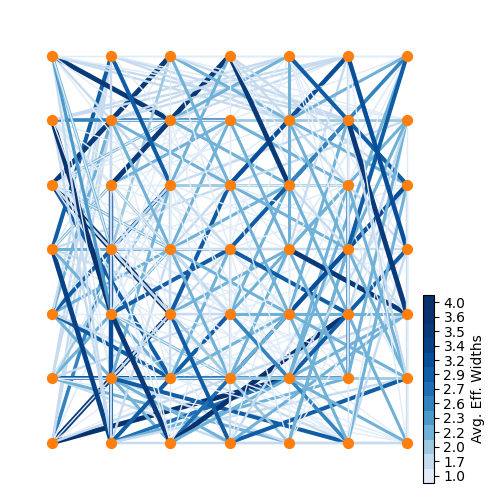}
   \caption{Simple bundles network ($L=3$) of a non-periodical orthogonal lattice with dimension $7 \times 7$, with both coordinates of the nodes perturbed by $\delta = 1e-12$. Despite these very small perturbations, a markedly heterogeneous and asymmetrical distribution of simple bundles can be observed. Interestingly, the perturbations allowed a substantial increase of the observed values of average effective widths.}\label{fig:Widths_lat_12}
\end{figure}

Figure~\ref{fig:Widths_lat_12} depicts the simple bundles network obtained for the network corresponding to the previous orthogonal lattice with both nodes coordinates perturbed by $\delta=1e-12$. After the perturbation, the nodes are interconnected in terms of the respectively established Delaunay triangulation.  More specifically, all pairs of nodes in the obtained triangulation are linked.  The visualization of the resulting network is shown in Figure~\ref{fig:nets}(a).

\begin{figure}
  \centering
     \includegraphics[width=0.7\textwidth]{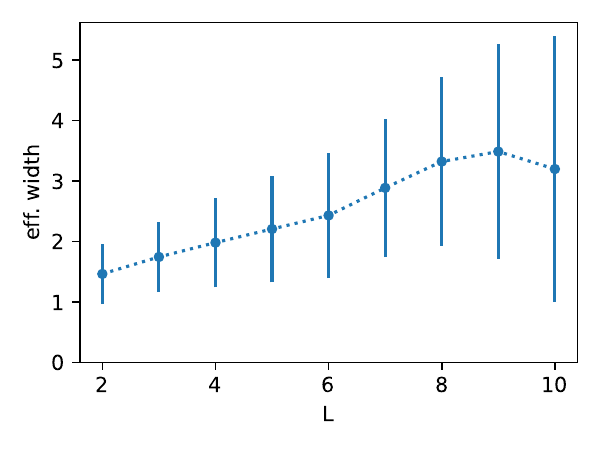}
   \caption{The average $\pm$ standard deviation of the effective widths obtained for the simple bundle networks in Fig~\ref{fig:Widths_L}. Interestingly, the standard deviation increases monotonically with $L$. This curve can be understood as a signature of the effective widths characterizing a given original network which, in the present case corresponds to the perfect non-periodical orthogonal lattice.}\label{fig:Signature}
\end{figure}

Figure~\ref{fig:Signature} depicts the average $\pm$ standard deviation values, in terms of the hierarchical level $L$, of the effective widths in the simple bundle networks shown in Figure~\ref{fig:Widths_L}, corresponding to a $7 \times 7$ perfect non-periodical orthogonal lattice.
It is of particular interest to observe that the average effective width reaches a peak at $L=9$, while the standard deviation values increase monotonically with $L$.

A substantially heterogeneous and asymmetrical network has been obtained even for these very small perturbations. Also interesting is the fact that substantially large average effective widths have been obtained. Additional experiments (not shown) indicate that the increase of average effective widths tends to take place involving the nodes at the border of the network.

\begin{figure}
  \centering
    \includegraphics[width=0.45\textwidth]{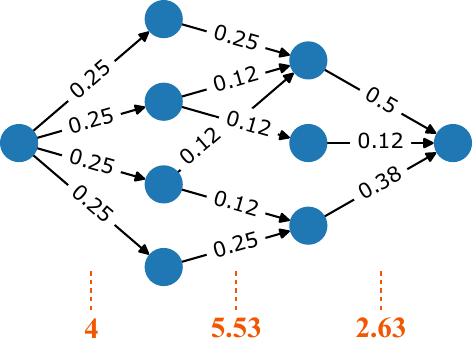} \hspace{0.5cm}
    \includegraphics[width=0.45\textwidth]{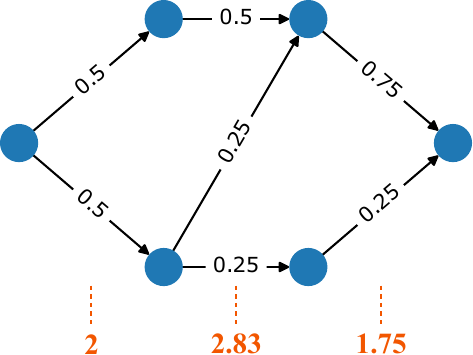}
    \\ \vspace{.5cm} (a) \hspace{.4\textwidth} (b)
    
    \caption{(a): One of the simple bundles obtained in the network shown in Fig.~\ref{fig:Widths_lat_12}. This specific bundle has widths $4.0, 5.66,$ and $2.65$ at the hierarchical levels 1, 2, and 3, yielding an average effective width of 4.1. (b): One of the simple bundles having maximum average effective width obtained for the case of the perfect non-periodical orthogonal lattice.}\label{fig:ex_bundle}
\end{figure}

Figure~\ref{fig:ex_bundle}(a) illustrates one of the simple paths obtained in the example above, which resulted with a relatively high average effective width equal to 4.1. The widths obtained along the three subsequent neighborhoods were $4.0, 5.66,$ and $2.65$. This result indicates that, respectively to the original perfect lattice, the average effective width tends to increase as a consequence of particularly large values of effective widths resulting at the intermediate hierarchical levels (5.66 in the case of the particular example above). These results provide an interesting example of a situation in which the incorporation of some disorder greatly contributes to changing some important properties (effective widths) of a network.

For the sake of comparison, Figure~\ref{fig:ex_bundle}(b) presents one of the simple bundle having maximum average width in the case of the perfect non-periodical orthogonal lattices.  Smaller effective widths can be observed along each of the three successive hierarchical neighborhoods.

The simple bundle in Figure~\ref{fig:ex_bundle} indicates that, despite the relatively high average effective width, the simple bundle actually has one level with a width of only $2.65$. Though still larger than the maximum average effective width in the perfect lattice, this minimum value still ends up limiting the flow in all other layers, deserving special attention.  This important fact motivates the analysis of simple bundles also to include the consideration of the minimum effective widths along the successive hierarchical neighborhoods, which is addressed at the end of this section.

\begin{figure}
  \centering
     \includegraphics[width=0.7\textwidth]{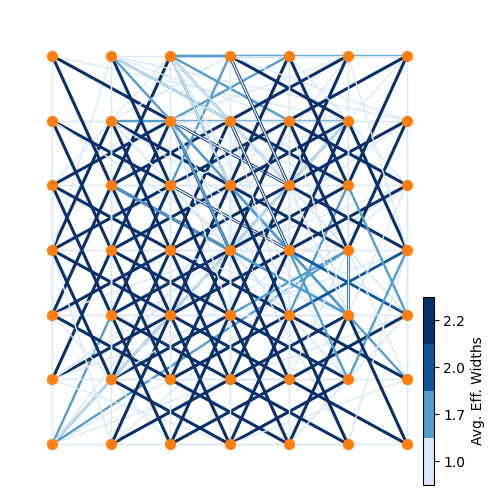}
   \caption{Simple bundles network ($L=3$)  of a non-periodical orthogonal lattice with dimension $7 \times 7$, obtained by using the Watts-Strogatz model with $p=0.05$.  Though only two rewirings were implied in this case, the obtained network resulted markedly distinct from the case involving a perfect lattice.  As with the spatial perturbations case, the topological perturbation implied an increase of the values of average effective widths.}\label{fig:Widths_latWS}
\end{figure}

The effects of topological perturbations to a non-periodical orthogonal lattice with dimension $7 \times 7$ are shown in Figure~\ref{fig:Widths_latWS}. More specifically, a respective Watts-Strogatz network was obtained with reconnection probability $p=0.05$. Figure~\ref{fig:nets}(b) presents the visualization of the so-obtained network. Similarly to the previous spatial perturbations, the topological modifications also tended to increase the average effective widths of the simple bundles network, though in a more general manner.   

\begin{figure}
  \centering
     \includegraphics[width=0.7\textwidth]{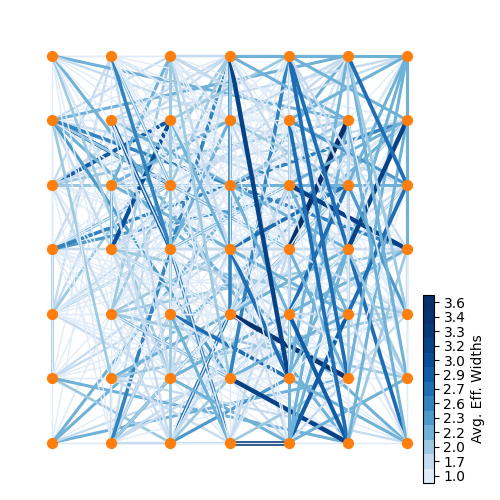}
   \caption{Simple bundles network ($L=3$)  of a periodical (thoroidal) orthogonal lattice with dimension $7 \times 7$, obtained by using the Watts-Strogatz model.  Though the same probability  $p=0.05$ has been considered in this case, the obtained simple bundles network resulted substantially different from the case involving the perfect lattice. }\label{fig:Widths_ws}
\end{figure}

The effects of topological perturbations to a periodical (thoroidal) orthogonal lattice with dimension $7 \times 7$ are shown in Figure~\ref{fig:Widths_ws}. Though the same reconnection probability $p=0.05$ has been adopted in this case, an even more distinct simple bundles network has been respectively obtained, implying that the periodical boundary contributed substantially to propagating the rewiring effects on the simple bundles throughout the network.

We now proceed to investigate the simple bundles networks obtained for the same three situations as above, but now taking into account the minimum effective width along the successive hierarchical neighborhoods instead of the average effective widths.

\begin{figure}
  \centering
     \includegraphics[width=0.7\textwidth]{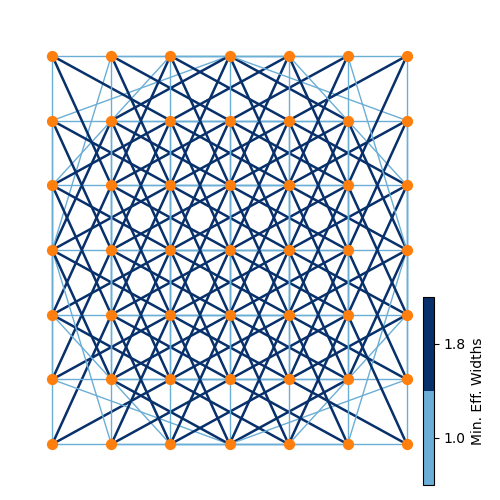}
   \caption{Simple bundles network ($L=3$) of the perfect non-periodical $7 \times 7$ orthogonal lattice considering the minimum values of the effective widths along the successive hierarchical neighborhoods.}\label{fig:Widths_lat_min_0}
\end{figure}

\begin{figure}
  \centering
     \includegraphics[width=0.7\textwidth]{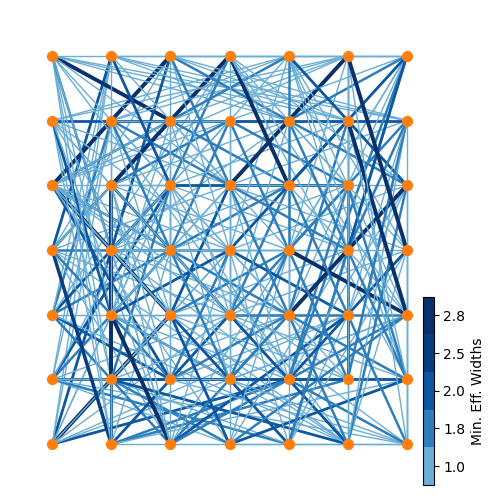}
   \caption{Simple bundles network ($L=3$) of geometrically modified non-periodical $7 \times 7$ orthogonal lattice considering the minimum values of the effective widths along the successive hierarchical neighborhoods.}\label{fig:Widths_lat_min_12}
\end{figure}

\begin{figure}
  \centering
     \includegraphics[width=0.7\textwidth]{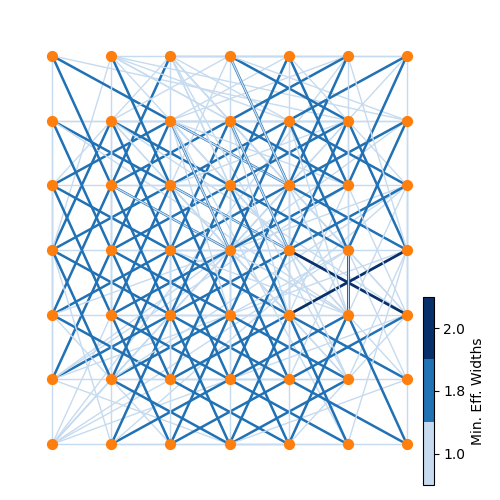}
   \caption{Simple bundles network ($L=3$) of topologically modified periodical (thoroidal) $7 \times 7$ orthogonal lattice considering the minimum values of the effective widths along the successive hierarchical neighborhoods.}\label{fig:Widths_min_latWS}
\end{figure}

\begin{figure}
  \centering
     \includegraphics[width=0.7\textwidth]{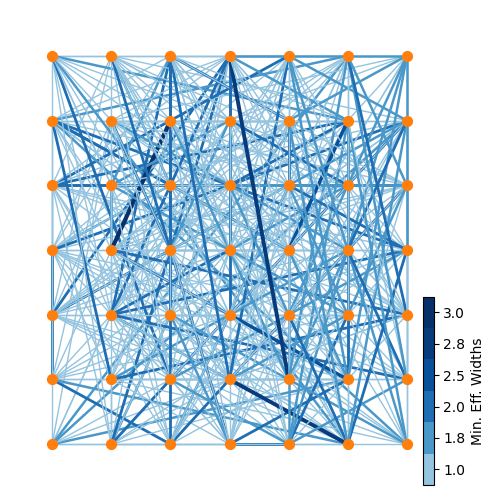}
   \caption{Simple bundles network ($L=3$) obtained for the Watts-Strogatz network considering the minimum values of the effective widths along the successive hierarchical neighborhoods. }\label{fig:Widths_ws_min_12}
\end{figure}

Figures~\ref{fig:Widths_lat_min_0},~\ref{fig:Widths_lat_min_12},~\ref{fig:Widths_min_latWS}  and ~\ref{fig:Widths_ws_min_12}  depict, respectively, the minimum effective bundles networks obtained for the perfect non-periodical lattice, the same with geometrical perturbations, and the non-periodical and periodical (thoroidal) Watts-Strogatz networks.  The resulting visualizations are almost identical to those obtained previously, except for the smaller weight values corresponding to the minimum effective widths. This result indicates that the average and minimum effective widths observed for most situations tend to be strongly positively correlated.

Though all the previous examples in this section considered the length of the simple bundles as $L=3$, it is also interesting to study how the simple bundles starting at a given source node change in terms of $L$. This topic is briefly developed in the present section.

\begin{figure}
  \centering
     \includegraphics[width=0.75\textwidth]{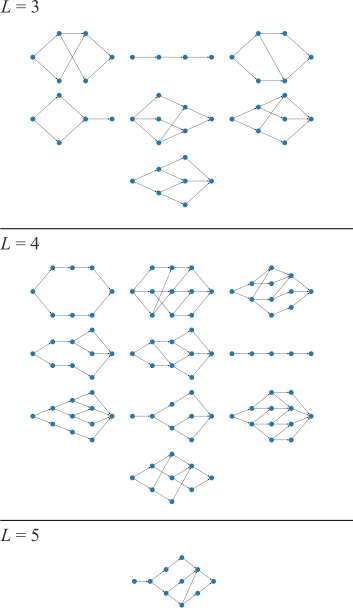}
   \caption{The distinct types of simple bundles defined while taking the central node in Fig.~\ref{fig:Widths_lat_12} as source for $L=3, 4,$ and $5$. The obtained simple bundles resulted markedly distinct one another as far as their respective topology is concerned.
    In all cases, the source node has been placed at the left-hand side.}\label{fig:L_bundles}
\end{figure}

Figure~\ref{fig:L_bundles} illustrates the simple bundles identified between the central node in Figure~\ref{fig:Widths_lat_12}, taken as source, and all possible destination nodes for successive values of $L$.

\section{Concluding Remarks}

Complex structures and systems are characterized by heterogeneous, diverse properties. A good deal of the research efforts in the area of network science have been invested in devising concepts and approaches that can be employed to obtain complementary and more comprehensive characterization of the topological properties of networks at several scales.

The present work addressed the possibility, initially described in~\cite{da2023quantifying}, of characterizing complex networks in terms of the properties of simple bundles defined between their pairs of nodes. More specifically, given a network and a pair of its nodes (source and destination), a methodology has been described that can be used to identify the set of simple paths that are successively followed as one moves along successive hierarchical neighborhoods referring to the source node, until reaching the respective destination node. By `simple', it is meant that the bundles never repeat the same node along distinct hierarchical neighborhoods.

In addition to identifying simple bundles, their effective widths along the hierarchical neighborhoods, which can be summarized in terms of the respective average, have also been defined in terms of the exponential entropy of the transition probabilities (or flow) throughout the successive neighborhoods.

The potential of the described concepts and methods has been illustrated respectively in the analysis of perfect as well as geometrically and topological modified orthogonal lattices. Several interesting results have been identified.  First, we have the identification of a few possible averaged effective widths in the case of unperturbed non-periodical lattices.  At the same time, even markedly small spatial perturbations have been verified to have a strong effect on the obtained simple paths.  This type of perturbation has been found, in the case of the considered networks, to lead to substantially larger average effective widths, typically involving nodes at the border of the original network.  The topological modifications implied by the Watts-Strogatz onto the considered periodical orthogonal lattice have been found also to yield substantially larger average effective widths throughout the network.

The effect of the parameter $L$ on the simple bundles obtained respectively to a given source node and all possible destination nodes has also been briefly addressed.  Quite distinct simple bundles have been obtained for $L=3, 4,$ and 5 while taking the central node of a spatially perturbed non-periodical orthogonal lattice.

The reported results pave the way for a number of further studies. For instance, it would be interesting to consider real-world structures, especially those possessing a geometrical nature, such as city and highway networks. Other interesting possible studies include the consideration of other network sizes and average degrees, as well as several lengths of simple bundles. It would also be interesting to consider progressive values of spatial and topological perturbations. Yet another possibility deserving further attention would be to study the simple bundles structure of hexagonal lattices.

\section*{Acknowledgments}
Alexandre Benatti thanks MCTI PPI-SOFTEX (TIC 13 DOU 01245.010222/2022-44).
Luciano da F. Costa thanks CNPq (grant no. 307085/2018-0) and FAPESP (grants 15/22308-2 and 2022/15304-4).

\bibliography{ref}
\bibliographystyle{unsrt}

\end{document}